# Mismatch Negativity: time for deconstruction

*F. Lecaignard and J. Mattout*

**Abstract -**

Error signals are the cornerstone of predictive coding and are widely considered essential to sensory perception and beyond. The mismatch negativity (MMN) is arguably the most emblematic and most studied brain error signal. It is affected in many brain disorders. However, its precise algorithmic function and the underlying physiology remain mysterious. Over the past decade, theoretical and computational explanations have been put forward. They highlight a paradox: the MMN is considered a signature of context-dependent perceptual learning, although it is defined as an evoked response averaged across trials, thus neglecting the information carried by error signal fluctuations over time. We propose to deconstruct the MMN, by virtue of hypothesis driven computational approaches whose aim it to account for these fluctuations.

## Introduction

Think of your favorite movie. It is made of single frames which, put together, contributes to a meaningful animation. Now, what would the average image from this film look like? Surprisingly, the result is not uniformly gray nor randomly pixelated but fairly presents a chromatic signature (Figure 1). One could expect extracting fine features to uniquely identify the film and assess differences with other films.

But the information needed to infer about the whole story is lost.

This experience is worth a thousand words to illustrate our claim where movie frames are replaced by brain signals measured over the course of a perceptual experiment. Precisely, we consider that predictive coding, a highly influential hypothesis that links these measurements to error signals, is today's best strategy to refine our understanding of sensory processing. Here, we argue that we need dynamic approaches to fully exploit this mathematical framework rather than traditional average-based models.

_ _ FIGURE 1 _ _

This paper focuses on one particular but essential auditory brain marker (the Mismatch Negativity, MMN; Figure 2), an electrophysiological response that quickly became emblematic of a prediction error calculated by the predictive brain (Heilbron and Chait, 2018). And we will rely on MMN models, here treated as mathematical objects. In empirical science addressing complex systems (such as the brain), models are necessary to understand hidden processes and simplifications are required to ignore out-of-scope mechanisms (Levenstein et al., 2023). In cognitive neuroscience, electro- and magneto-encephalography (EEG, MEG) are two powerful techniques, non-invasive and millisecond-resolved, to measure brain activity at the surface of the scalp. One dominant model providing the MMN is the event-related potential/field (ERP/F) obtained by averaging multiple responses (called *single trials*) induced by the same stimuli (Figure 2.c). Thanks to trial averaging, ERP/F are assumed to be cleaned from undesirable inter-trial fluctuations (Picton, 1995), here referred to as the *experimental* noise.

_ _ FIGURE 2 _ _

But what if this noise become meaningful? What if each trial could be interpreted just like each frame of a movie is? And what if transitions between trials could reveal the whole story of the brain adaptation to its sensory environment?

The past two decades have seen the development of computational theories of brain functions, inspired by mathematical frameworks and the view of the brain as an information processing system (Dayan et al., 1995; Friston, 2012; Mumford, 1992). These disrupting approaches propose a *predictive brain* endowed with learning abilities to keep tracking the ever-changing regularities of the world. Addressing the learning entertained by the brain offers for the first time a means to test quantitatively the informational value of inter-trial fluctuations. The brain is indeed described as a system whose hidden states represent mathematical quantities (e.g. hierarchical prediction errors). These are updated each time a sensory input is processed and the related changes should therefore be reflected in EEG/MEG activity on a single-trial basis- and likely cancelled out through averaging.

This highly innovative context leads us to here question the future of the MMN, obtained from ERP/F. The MMN remains unchallenged as an informative and clinically useful neurophysiological marker of healthy and pathological perceptual processing (Fitzgerald and Todd, 2020; Näätänen et al., 2007). It was also essential to the widely accepted assumption that the processing of sensory information in human and animal brains follows a general principle based on error computation (Winkler, 2007). However the MMN faces a major limitation, and not the least: we still don't understand its functional role or physiological origin (Garrido et al., 2009b).

In this paper, we argue that the MMN model has become too simple to further characterize error processing and improve perception research. This speaks to examining the trajectory of brain responses over the successive stimulations. We advocate a switch towards dynamic modeling leveraging on recent computational theories of sensory processing combined with EEG/MEG analysis on a trial-by-trial basis.

We briefly present the MMN duality combining undebated contributions to perception research and critical unanswered questions. We then introduced to predictive coding, a learning framework providing a very influential account of the MMN. In the light of recent findings challenging its explanatory power, we claim the importance of taking a mechanistic approach as afforded by trial-by-trial modeling. We finally discuss initial modeling findings that offer promising perspectives, in clinical research in particular.

**Success and limitations of the MMN**

As illustrated in Figure 2, the MMN is obtained easily and rapidly, with simple instructions and no need for an explicit attentional engagement of participant. It typically involves an *oddball* paradigm which comprises sequences of repeating stimuli (standards) and rare unexpected ones (deviants). Using the ERP/F approach, the MMN is observed by subtracting the response induced by standards from that induced by deviants (Näätänen et al., 1978). Easy and successful, it has motivated an incredibly wide range of studies, with various original designs and numerous clinical applications (Michie et al., 2016; Näätänen, 2003; Sussman et al., 2014). The MMN appears ubiquitous (it is measured in several sensory modalities, several states of consciousness, in animals, from newborns to the elderly; Fitzgerald and Todd, 2020; Näätänen et al., 2011, 2007) and could therefore reflect a general principle of sensory processing based on error detection. Moreover, it is also observed in the case of more sophisticated statistical rules than sound repetition, combining one or multiple sensory modalities or in language and music (Näätänen et al., 2010). Overall, there is a large consensus that MMN is one of the most robust and universal markers of sensory and even cognitive adaptation (Näätänen et al., 2001).

What is the role and neural implementation of the MMN? The memory-based model (Naatanen et al., 2005; Winkler, 2007) describes the MMN as an error signal resulting from the comparison of the incoming sensory input with internal representations in a sensory memory. The adaptation model (Fishman, 2014; May and Tiitinen, 2010) explains the MMN in terms of a differential neural habituation between repeated standard and deviant auditory N1 components. A third view, the predictive coding model (PC, see next section) derives from the above-mentioned general *predictive brain* framework and accounts the MMN as a context-sensitive prediction error (Friston, 2005).

To assess which model best explains the MMN, model comparison has been addressed in several ways: using simulations (Lieder et al., 2013b; May, 2021; Wacongne et al., 2012), experimental manipulations (Parras et al., 2017; Schröger and

Wolff, 1996) or neural implementation (Garrido et al., 2009a; Lecaignard et al., 2021). Despite these efforts, a puzzling feature of the MMN is that we still fail to have a clear understanding of what it represents. Notably, this quest remains very active (Carbajal and Malmierca, 2018; Denham and Winkler, 2020; Fitzgerald and Todd, 2020; Garrido et al., 2009b; Heilbron and Chait, 2018; May, 2021); as a matter of fact this demonstrates the great potential of finely characterizing error processing to foster perceptual research.

In the following, we argue that there is an alternative way than the MMN to address error processing; it implies testing quantitatively how the brain learns sensory regularities (without which no error can emerge).

**Predictive Coding provides a learning-based account of the MMN**

Beyond the scope of perception, PC is first and foremost a generic algorithm for regularity learning based on minimization of errors (Spratling, 2017). As shown in Figure 3.a, it is based on a recurrent message passing in a hierarchical network, where each node calculates the difference between bottom-up observations and top-down predictions, yielding an error signal called a prediction error (PE). Hierarchical PEs trigger the revision of their related predictions until being explained away. Different learning styles ($g, f$ functions in Figure 3.a) can be implemented in PC including Bayesian inference (Figure 3.b). Information processing in a bayesian framework accounts for the reliability of information (called *precision*) which depends on the context. Regularity learning operates on a trial-by-trial basis and its efficiency derives from its speed (or learning rate) indexed on the contextual relevance of PEs (Mathys et al., 2014). Context-sensitive errors are called *precision-weighted prediction errors*.

_ _ FIGURE 3 _ _

PC could be implemented by any computer or agent including the brain as assumed under the *predictive brain* (Friston, 2005). Oddball paradigms are extremely convenient to test this hypothesis: standards and deviants enable assessing regularity learning and adaptation to change through error signaling, respectively. The MMN is seen as a precision-weighted prediction error (Figure 4), namely an error term that coincides with former models of the MMN [1] (Carbajal and Malmierca, 2018; Garrido et al., 2009b) multiplied by a precision gain. Interestingly, contrary to the memory-based and the adaptation models (but see May, 2021), PC is not restricted to deviance processing but addresses the processing of every input of the oddball stream. Its hierarchical structure enables investigating the learning of all kinds of regularities, from sound repetition to long temporal associations (Kanai et al., 2015; Kiebel et al., 2008).

This PC model is a real breakthrough as it is based on the brain solving probabilistic inference (Pouget et al., 2013). Its validity was first established from simulation studies reproducing known modulations (Lieder et al., 2013b) and biological plausibility could be further supported (Garrido et al., 2009a). It quickly became influential not only because of its great explanatory power (Heilbron and Chait, 2018; Winkler and Czigler, 2012; Winkler and Schröger, 2015) but also by highlighting the dynamical nature of perception (De Lange et al., 2018). Considering that information processing is facilitated in predictable (or temporally structured) environments, several studies addressed the MMN modulation by stimulus predictability (Figure 4.c), although this hypothesis was long assumed incompatible with the automaticity of the MMN (Scherg et al., 1989).

PC is arguably the most influent MMN model today. It has largely contributed to examine sensory processing from a novel perspective that is not restricted to deviance processing and emphasizes the context sensitivity of perception at multiple timescales.

**A model both too complex and too simple?!**

However, some criticisms could emerge recently, two of them in particular illustrate our claim for taking a mechanistic turn.

*Too complex?* This points refers to the putative complexity of PC and to the long-studied comparison between the PC and the adaptation models (Garrido et al., 2009a).

---

[1] *Rigorously speaking, the MMN should be interpreted in PC as the difference between the deviant and standard errors.*

The adaptation model appears a fairly well-understood cortical mechanism, involving the decrease of post-synaptic responsiveness by redundant stimulations (May and Tiitinen, 2010). At the cellular level in animals, oddball stimulations show a similar deviance-based response (SSA; Ulanovsky et al., 2003). Mismatch responses measured at multiple spatial scales thus suggest adaptation to be a fundamental brain principle, described as "simpler" than PC (Carbajal and Malmierca, 2018; Heilbron and Chait, 2018), "low-level" (Fitzgerald and Todd, 2020), "stimulus-driven" (Heilbron and Chait, 2018), "bottom-up" (May, 2021) or "local" (SanMiguel et al., 2021). On the contrary, PC is presented as an active process requiring top-down information from high-level regions (Heilbron and Chait, 2018; Strauss et al., 2015). Applying the principle of Ockham's razor, the adaptation model should be selected over PC in the case of typical oddball processing (Heilbron and Chait, 2018; May and Tiitinen, 2010; O'Reilly and O'Reilly, 2021).

However, May recently acknowledged that "*there is nothing low-level about adaptation*" (May, 2021). Furthermore, if one make a clear distinction between a psychological and a physiological description of information processing following D. Marr's vision (Marr, David, 1982), the PC and the adaptation models appear as the two sides of the same coin, mental processes and biological implementation, as already suggested by some (Stefanics et al., 2016). Because of their different levels of analysis they cannot be rigorously compared (Levenstein et al., 2023). That said, it is crucial to address both for an exhaustive description of sensory processing.

*Too simple?* It is suggested by recent unexpected findings as they challenge the predictive power of PC. Chait and colleagues expected the reduction of MEG power during the listening of auditory streams alternating sounds according to predictable transitions as compared to randomly but they measured the opposite effect (Barascud et al., 2014; Sohoglu and Chait, 2016; Southwell et al., 2017). Todd and colleagues tested the hypothesis of a larger MMN in more stable contexts using oddball sound sequences and they unexpectedly observed a block-order effect affecting sustainably standard tone representation (Todd et al., 2013b, 2014). In both cases, the authors pointed to the hitherto underestimated role of precision weights, that amplifies contextually relevant PEs while silencing uninformative ones (Barascud et al., 2014; Fitzgerald and Todd, 2020; Heilbron and Chait, 2018; Todd et al., 2013b). Figure 4.a,b provide examples of MMN modulations with varying precision weights and PE representations by the brain; they illustrate the non-bijectivity of the MMN-PC mapping. Notably, in Lecaignard et al. (2022a), we explained the opposite effect of predictability on PE and precision weights: structured sequences yield less surprising deviants (reduced PEs) and convey more information (increased precision weight). The two quantities become inseparable when multiplicated to derive a precision-weighted prediction error (Figure 4.c).

_ _ FIGURE 4 _ _

We argue that the two criticisms vanish as we work with trial-by-trial time series (Figure 5) to seek evidence for an explicitly-described learning in electrophysiological signals. These evidences pertain precisely to the dynamics of learning quantities (PEs, precision weights,..) that we lose with ERP/ERF averaging (Figure 6).

**Dynamic modeling of oddball responses.**

The brain as a Bayesian machine remains an open-question and the combination of PC with single-trial modeling appears as today's best strategy to enrich perceptual models. Important, this turn does not imply starting this research *de novo* but rather builds on the fruitful MMN research in the past forty years.

*Practical guidelines.* Figure 5 describes how trial-by-trial timeseries are extracted. Dynamic modeling analysis is performed separately at multiple spatio-temporal data points defined for every (virtual) sensor and peri-stimulus temporal samples of interest, an aspect that gives rise to intensive calculations. An entire procedure is proposed in Lecaignard et al. (2022a, 2022b). First, the model space encompassing alternative hypotheses should be defined (in particular, functions $g, f$ in Figure 3.a should be defined for each model, as well as the mapping of model quantities onto brain signals). Figure 6 shows learning and non-learning error trajectories that can be confronted to the real EEG data. Then, model inversion for each subject, each model, each data point should be run. Statistical analysis leverages on bayesian model comparison to select the most plausible model at the group-level. We obtain a spatio-temporal description of the process(es) entertained by the brain during the experiment.

Open-source toolboxes with very helpful documentation enable a fairly accessible implementation (Daunizeau et al., 2014; Frässle et al., 2021). We also strongly encourage reading Stephan's recommendations for model inversion and group-level statistical analysis in a Bayesian framework (Stephan et al., 2010).

_ _ FIGURE 5_ _

Contrary to conventional MMN studies, this model-driven approach is highly dependent on model space and experimental design. Its key advantage is that variables of interest become directly interpretable because they reflect mathematical quantities like a learning rate or prediction update; interpreting physiological measures like peak amplitude or latency is clearly not so straightforward. In addition, assuming a model-based experimental manipulation, learning quantities reveal context-dependent modulations that can be directly seek in brain data (Figure 4.a).

*Scarce but robust evidence.* We here present a short review of the emerging literature with scarce but consistent oddball findings supporting inter-trial EEG/MEG fluctuations as evidence for Bayesian learning.
An early and influential EEG study looked at the evolution of the deviant peak over the experimental time, extracted in the MMN time window within a fronto-central cluster (Lieder et al., 2013a). Group-level model comparison showed that Bayesian learning best explained these data. In other EEG/MEG studies involving different sensory modalities and examining brain data comprehensively, learning models were found to outperform static models consistently across studies, within spatio-temporal clusters that show a significant mismatch effect using a classical ERP approach (Lecaignard et al., 2022a; Ostwald et al., 2012; Poublan-Couzardot et al., 2022; Stefanics et al., 2018). Consistency between ERP and single-trial modeling could not be fully observed in a study addressing the effect of ketamine on oddball responses (Schmidt et al., 2013; Weber et al., n.d.). This highlights the different interpretations that can be drawn from a partial or complete (dynamic) analysis – as illustrated by the movie example presented in the introduction.
Predictability manipulation could reveal a context-sensitive time-dependency during oddball processing (Lecaignard et al., 2022a) and different learning styles were found at different latencies of auditory processing in (Maheu et al., 2019) . Also, sequences of sounds sampled from gaussian distributions enabled to characterize the mapping of computational quantities onto frequency bands, using depth electrodes (Sedley et al., 2016).

_ _ FIGURE 6 _ _

**Perspectives**

The trial-by-trial modeling of brain responses, here presented in the context of oddball studies, enables selecting the most plausible generative process of the overall dataset. The winning model encompasses a mapping of model states onto electrophysiological features and thereby provides a computational interpretation of each single-trial response. Dynamic modeling thus paves the way for the functional interpretation of brain activity *on the fly*. If this sounded like science fiction a few years ago, it now seems reachable - even if a number of challenges still need to be overcome (see below).

This offers exciting opportunities in the field of on-line analysis of brain activity. This technique is central to brain-computer interfaces and neurofeedback protocols both expected to face important clinical challenges such as communication restoration or attention training (Wyckoff and Birbaumer, 2014). On-line analysis for clinical applications are mostly developed in EEG for convenience with patients. They are based on extracting and interpreting on the fly canonical ERPs such as the MMN or the P3 (Mattout et al., 2015). However, experience from our lab shows that these physiological features are not always recognizable on a single-trial basis despite improvements in on-line data cleaning (Freitas et al., 2020). Two complementary approaches appear really worth testing: the use adaptive design optimization, a model-based approach to generate personalized stimulation stream (Sanchez et al., 2016) and the proposed modeling framework to provide cognitive biomarkers related to trial-wise predictions. The recent development of new MEG sensors suitable for a clinical use at the patient's bedside (Gutteling et al., 2023) could even more contribute to this exciting perspective.

Working at the single-trial level points to the challenging question of the neurobiological origin of surface-based EEG-MEG signals. Interpreting electrophysiological data which sums spatially and temporally the contribution of many cerebral and non-cerebral sources is acknowledged to be a difficult question (Cohen, 2017). Recent advances in artificial neural networks (ANN) could help at tackling this challenge in trying to bridge macro-scale and micro-scale electrophysiology. Note that auditory oddball responses are good candidates to assess this mapping because, at the microscopic level in animal models, oddball sounds elicit the SSA along the auditory system (Carbajal and Malmierca, 2018). A remarkable aspect of both SSA and the MMN is that they remain poorly understood and an active research currently aims to characterize error processing along the auditory pathway by questioning the SSA-MMN relationship under predictive coding (Escera and Malmierca, 2014; Malmierca and Auksztulewicz, 2021; Parras et al., 2017)

Another promising perspective, at hand given the present argumentation, consists in developing a personalized care to psychiatric conditions. This approach is inspired by new hypotheses emerging from computational psychiatry extending the *predictive brain* to psychiatric disorders (Adams et al., 2013; Friston et al., 2014). In schizophrenia, there is robust evidence for a reduced MMN, making it relevant for testing a dysfunctional NMDAR system (Todd et al., 2013a). However, this model could reach a limit because the neurophysiology under the MMN reduction remains unclear (Michie et al., 2016). At the cognitive level, the PC account of the reduced MMN is extremely influential but predictability effects measured in patients show disparate modulations over studies (for review, Kirihara et al., 2020). The assumption of a systematic MMN reduction is currently challenged as perhaps too simplistic, calling for more subtle investigations to test the proposed impairment of precision weighting accounting for psychosis (Adams et al., 2013; Cassidy et al., 2018; Corlett et al., 2019; Weilnhammer et al., 2020).

In autism, the large number of MMN studies reveals above all a high variability of results and points to a lack of consensus on how MMN might differ (or not) in autistic versus neurotypical participants. Again, computational psychiatry promises to clarify this issue since, as described by Haker et al. (2016), it aims to describe quantitatively and at the individual level the multiple dysfunctions associated with autism. However, a recent review of first empirical attempts concludes that such research is difficult (Angeletos Chrysaitis and Seriès, 2023). Dynamic modelling of low-level auditory processing is hopefully a good start.

Of course, this emerging field of research faces limitations exist, some of them already reported. A major pitfall is about the poor performance of model fitting typically expressed by the percent of data variance explained by the model (R2). Whenever reported, maximum values are around 5% (Lecaignard et al., 2022b; Maheu et al., 2019; Poublan-Couzardot et al., 2022; Sedley et al., 2016). Reliability of findings is however supported by looking at the R2 timeseries along the peri-stimulus time (each model inversion provides a R2 value at each peri-stimulus sample, Figure X). In Maheu et al. (2019) and Lecaignard et al. (2022b), R2 is found peaking at time windows overlapping significant auditory evoked component and close to 0% otherwise, suggesting that informative time samples (at least for an evoked analysis) could be recognized by the model. Clearly, there is a need to gather more observations to increase the statistical sensitivity and test the consistency of findings across studies, be it through large groups of subjects or with highly-informed datasets like simultaneous EEG-MEG recordings. Also, characterizing the mapping of model quantities (e.g. prediction errors or learning rates) onto brain signals deserves a strong interest; fitting data features related to brain rhythms is clearly a perspective we envisage to test their possible role under a predictive coding account (Friston et al., 2015; Sedley et al., 2016). When addressing sensory processing without an explicit attentional engagement (as is typically the case in most MMN studies), the lack of behavioral data prevents from inferring which learning style the brain entertains (Stefanics et al., 2018; Weber et al., n.d.). This issue is typically overcome by testing alternative learning combined to alternative mappings, an approach that increases the size of model space with an inevitable cost in terms of computational resources. Improved fitting performances are also expected from finer learning algorithms that could account for multiple subtle contributions subsumed in the single-trial data ; these could reflect parallel processings at different timescales. As model space becomes able to capture subtleties in brain signals, model separation becomes more critical, requiring smart experimental protocols to disentangle model predictions.

Another important limitation comes from the fact that data is typically extracted at a specific sample of the peri-stimulus time to derive a trial-by-trial timeserie (Figure 5). Such a time-locked extraction ignores the likely inter-trial jitters of brain processing in the temporal dimension. This may degrade model fitting performances and creates a time-locked dependence of model comparison. Fortunately, this issue could be fixed rather easily thanks to the bayesian setting that is perfectly relevant to combine multiple data.

**Conclusion**

Building models of perception takes time and the MMN has successfully contributed to improve them over the past four decades. Thanks to MMN findings, and those recent obtained in the theoretical guidance of perceptual learning, the amount of knowledge about sensory processing has enabled better representations that now require mathematical (computational) models rather than phenomenological ones. Therefore, we believe that the MMN has become a too simple model (Figure 6.b), perpetuating a main focus on deviance processing. And we underscore the somehow inconsistency of testing learning-based hypotheses using the MMN, as the former rely on dynamic processes whereas the latter is obtained through averaging. In turn, we emphasize the great informational value of trial-by-trial updates of brain signals that learning models have made investigable for the first time. We claim that perception research will progress significantly only if, using dynamic modeling, we take the opportunity to tackle the cognitive meaning of these specific data. Only this way can we fully exploit the rich temporal information afforded by electrophysiology to characterize brain adaptation to external changes.

_ _FIGURES_ _

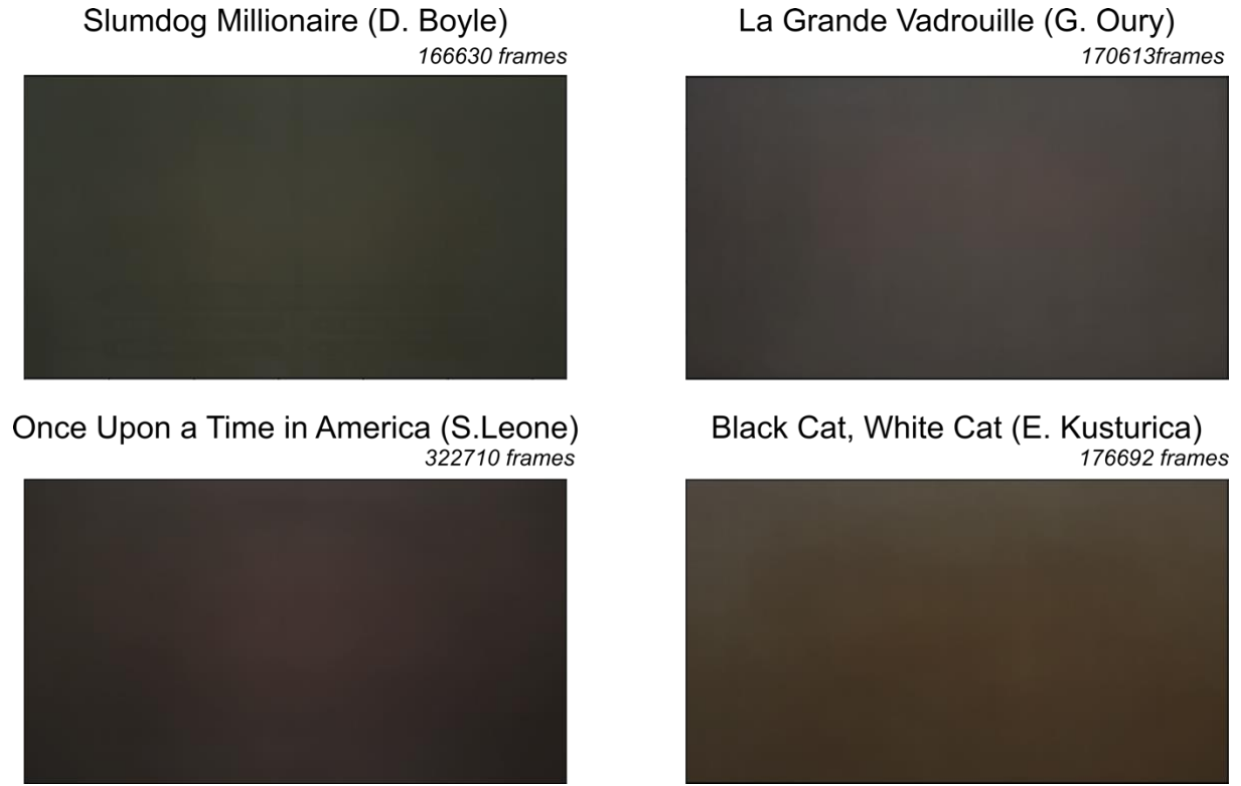

***Figure 1. Examples of RGB average frame.*** *For each example, average was computed with RGB values treated separately for each pixel. The first and last 5 minutes of frames were excluded to remove opening and closing credits. Average frames have been here scaled to the 16/9 format. Although the first impression might be uniform and similar renderings, a deeper examination reveals differences between image center and edges, and shade differences across movies (from top to bottom, left to right: greenish, pinkish, reddish, yellowish). These effects suggest that averaging retains information. Images were generated using a Python code inspired by ChatGPT.*

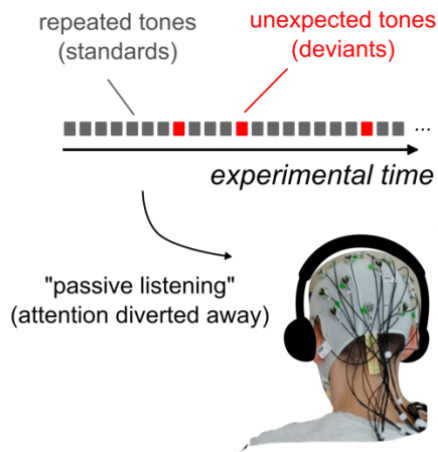
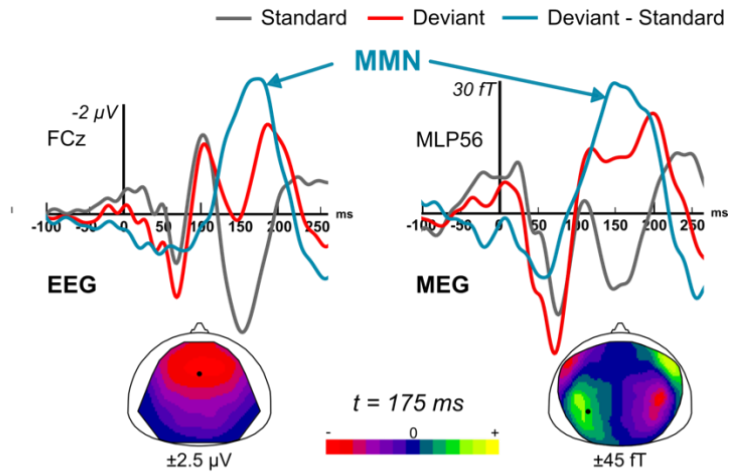
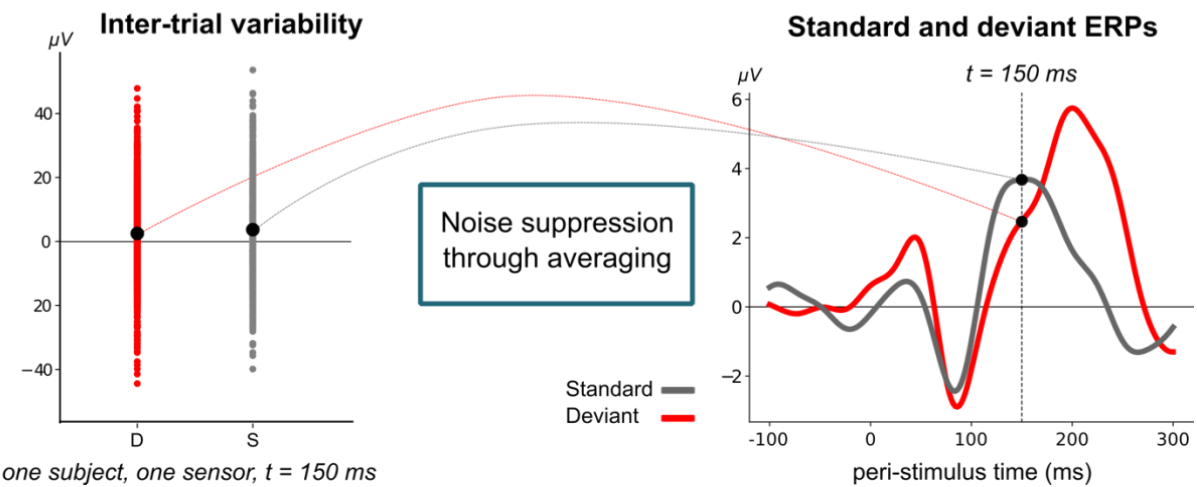

*Figure 2. **a) Typical oddball sound sequence**. Standard (grey) and unexpected deviant (red) tones are delivered to the participant while he/she is asked to not pay attention to the sounds. A typical EEG setting is shown. **b) Auditory evoked responses** to the standards (preceding a deviant; grey), deviants (red) and their difference (blue) measured using EEG (left; average-reference) at a frontal electrode (FCz) and MEG (right) at a left posterior temporal gradiometer (MLP56). Blue arrows point to the MMN component in each modality. Scalp topographies (bottom) obtained in each modality at latency t= 175ms with FCz (left) and MLP56 (right) locations (black dots). Amplitude range and color code are provided for each plot. EEG and MEG data are group averages described in Lecaignard et al. (2021). **c) Illustration of the ERP/ERF model principle**, using an individual EEG dataset measured in a typical oddball paradigm (>5000 trials, EEG data: nose-reference, 2-20 Hz band-pass filtered and baseline-corrected). Left: single-trial signal amplitudes measured at t=150ms at electrode Cz are sorted by deviant (D), standard preceding a deviant (S) categories. Black dots indicate average amplitudes per sound category. Right: the ERP model provides the averaged traces for standard (preceding a deviant, grey) and deviant (red). Black dots point to average amplitude at peri-stimulus latency t=150ms. Red and grey arrows link the averages obtained at t=150ms for deviant and standard tones, respectively, to highlight the same average amplitude between panels.*

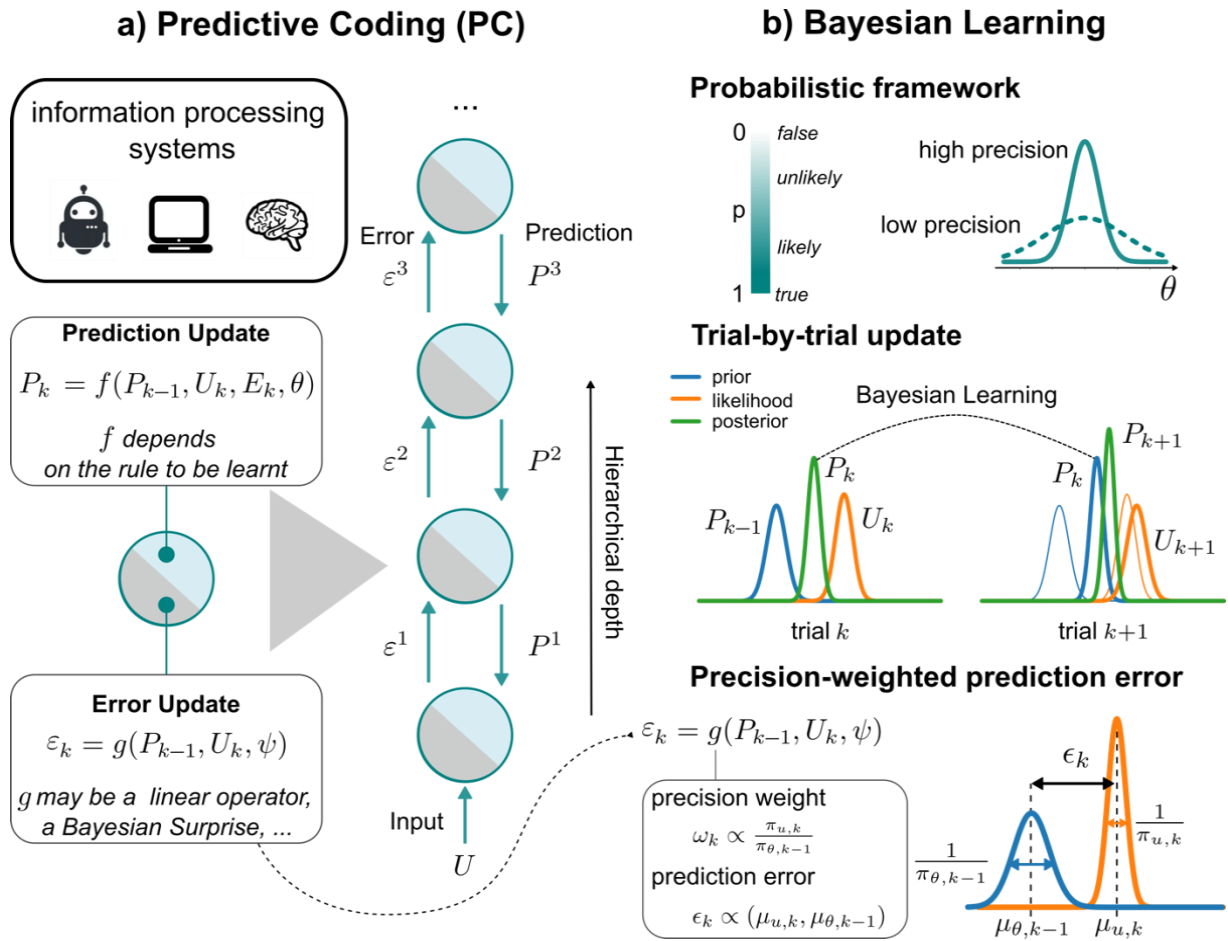

*Figure 3. Rationale. a) Predictive Coding. The general hierarchical structure with ascending order levels is depicted on the right. Between levels, ascending and descending connections convey prediction errors ($\varepsilon$) and prediction ($P$), respectively. Within-level computations (right) involve an error unit (grey) that generates a prediction error indexing a mismatch between the level's prediction and input. A prediction unit (light blue) enables updating prediction. Both error and prediction computations are explicitly defined by the modeler as functions ($g, f$ resp.) relying on past experience ($P_{k-1}$), current observation ($\varepsilon_k / U_k$) and some parameters ($\psi, \theta$). Top-Left insert highlights that algorithms ($g, f$) can be implemented by any systems processing information, including the brain. b) Bayesian Learning. Top row: in a Bayesian setting, probability $p$ refers to the plausibility of an information indexed between 0 (impossible) and 1 (certain).The variance of probabilistic distribution is the mathematical translation of information reliability, illustrated here by two gaussians.. Middle row: Principle of Bayesian learning. At trial $k$, the initial belief instantiated by the Bayesian agent (prior distribution, blue) is confronted to current observation (likelihood, orange). The resulting update (posterior, green) is obtained by applying Bayes's rule. Next, the posterior belief becomes the prior belief (blue) at trial $k + 1$. Previous round distributions (trial $k$) are displayed with thin lines. Bottom: Bayesian learning is implementable in PC yielding context-sensitive prediction errors. An example based on gaussian distributions is depicted. In this case, precision weight $\omega_k$ indexes the ratio of prior and likelihood precisions and the error $\in_k$ integrates the mismatch between current observation and prediction beliefs. Both quantities, $\omega_k$ and $\in_k$, evolve over trials as specified in the learning model ($g, f$).*

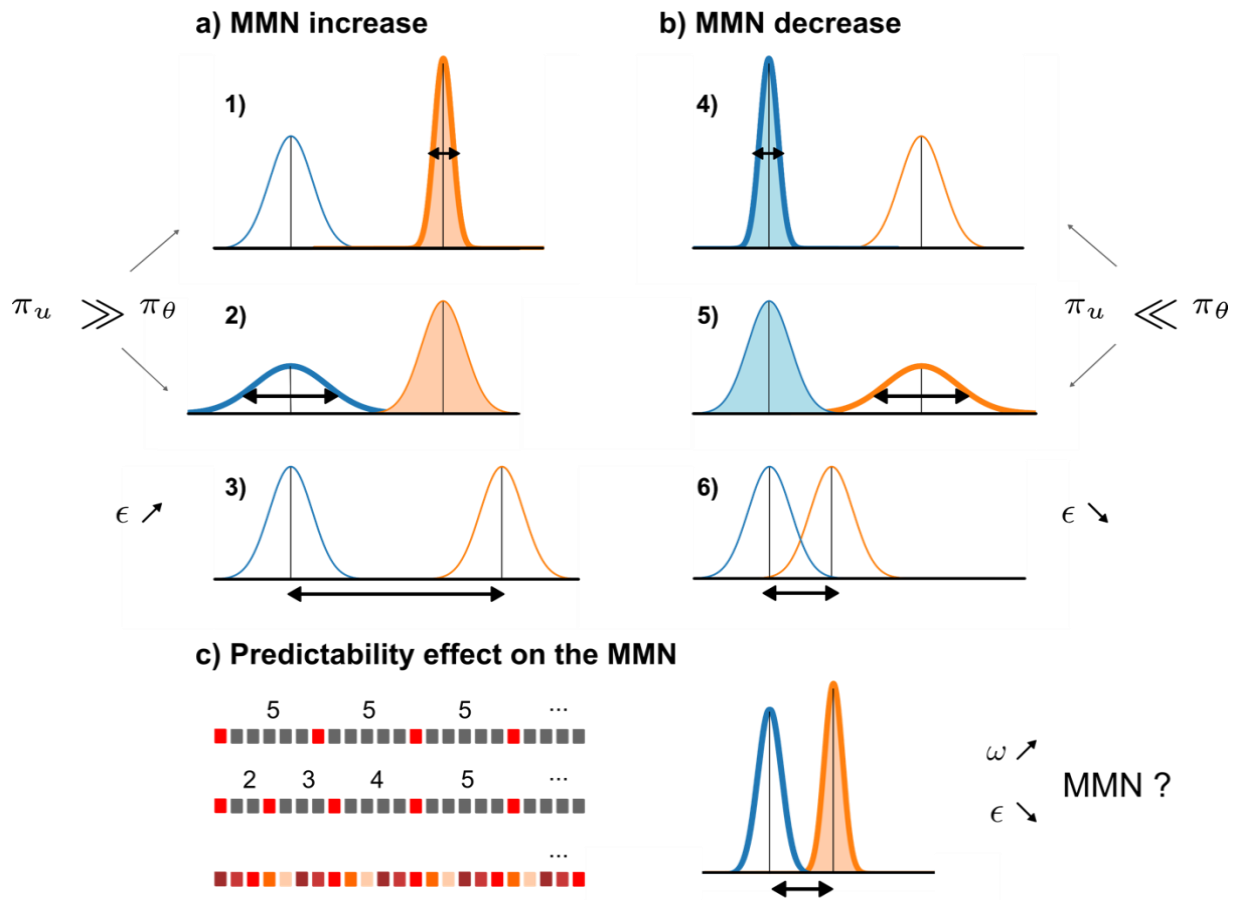

*Figure 4. **The PC model of the MMN predicts context-sensitive effects**. Top: under PC, the MMN is a precision weighted prediction error. Confrontation of current belief and incoming observation is illustrated using gaussian distributions (blue and orange traces, resp.). These two distributions then serve as baseline to illustrate 6 specific predictions of MMN modulation. **a) MMN increase** is expected in cases 1:3. Cases 1 and 2 show a larger learning rate (as compared to baseline, top panel) obtained through 1) a more precise likelihood (ex: high quality stimulation) or 2) a flat prior (ex: at the beginning of an oddball experiment when the participant knows nothing about the sequence). Case 3 depicts a larger standard-deviant magnitude. **b) MMN decrease** is expected in cases 4:6. Cases 4 and 5 show a lower learning rate due to 4) a more precise prediction (ex: long exposure to the repetition rule) or 5) an unprecise observation (ex: noisy acoustic environment). Case 6 depicts a lower standard-deviant magnitude. Colored distributions indicate the more precise information whenever the two distributions differ in variance. Notation subscripts related to trial have been removed for clarity. These examples (among many) aim to show that multiple explanations can produce similar modulation of the MMN, being a global learning index measured by the experimenter. **c) Predictability effect**. Left: three examples of predictable rules are presented inspired by Dürschmid et al. (2016), Lecaignard et al. (2015) and Barascud et al (2014). Grey/red code for standard/deviant; from red to light orange for different sounds*

organized into repeated streaks. Numbers indicate the number of repeated standards. Right: as compare to baseline (typical oddball rule, Figure 2), predictable deviants are expected to reduce the error term ∈ and increase the precision weight ω (arguably by increasing the sensory gain towards structured sound sequence, Lecaignard et al., 2022a). Consequently, the predictability effect on the MMN cannot be predicted unless the precise dynamics of ∈ and ω is examined.

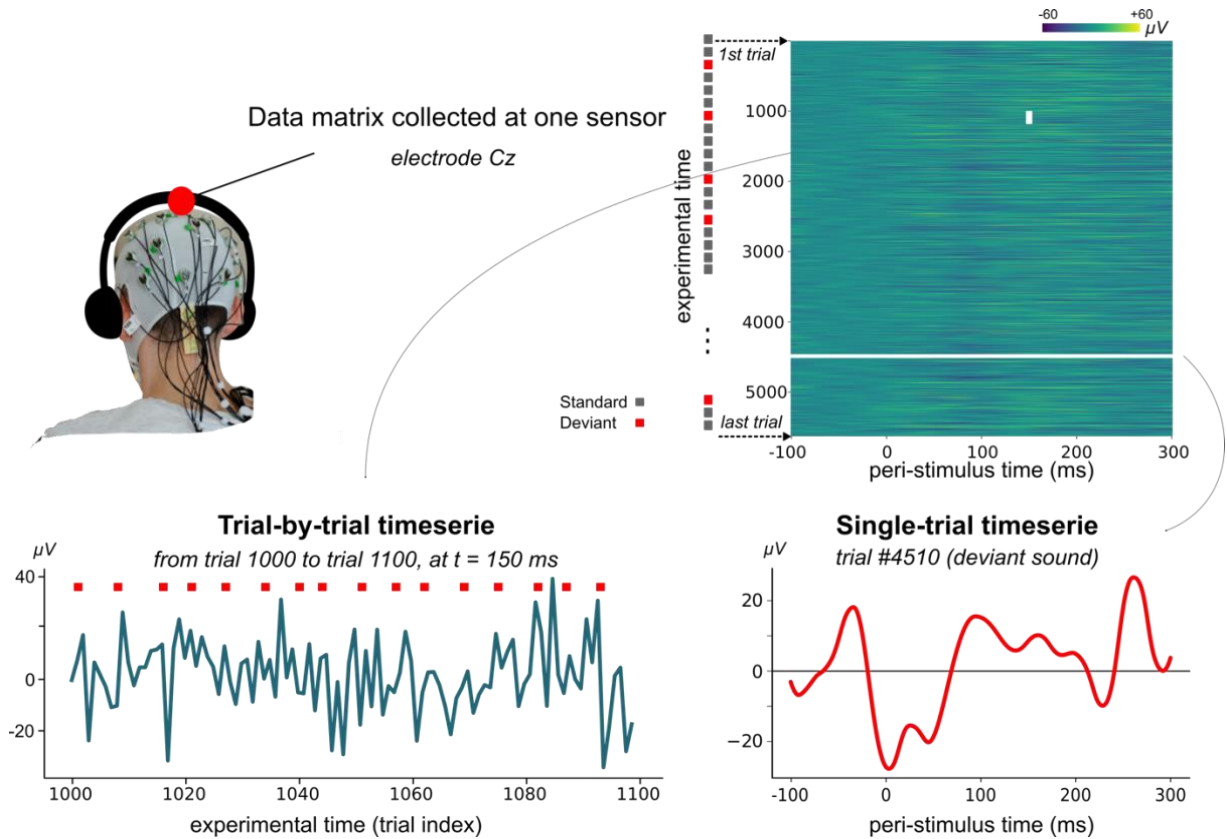

*Figure 5. Two types of temporal timeseries extractable from electrophysiological data.* Considering an dataset collected in a single participant, data at one particular sensor (here oddball EEG data at electrode Cz), can be represented as a 2d matrix where rows are the single responses sorted by experimental order (from up to bottom; >5000 trials in the present example) and columns span the peri-stimulus time samples (here 200, from -100 to 300 ms). The vertical white rectangle highlights the data measured at Cz, at peri-stimulus time t=150ms from trial 1000 to trial 1100. The related timeserie is depicted (bottom-left panel) with deviant stimuli (red squares) overlay. The horizontal white rectangle at trial 4510 highlights a single-trial responses, shown on the bottom-right panel. EEG data is 2-20 Hz band-pass filtered, nose-referenced and baseline-corrected. Amplitude signal is indicated by color-scale.

## a) Dynamic PC model of oddball processing

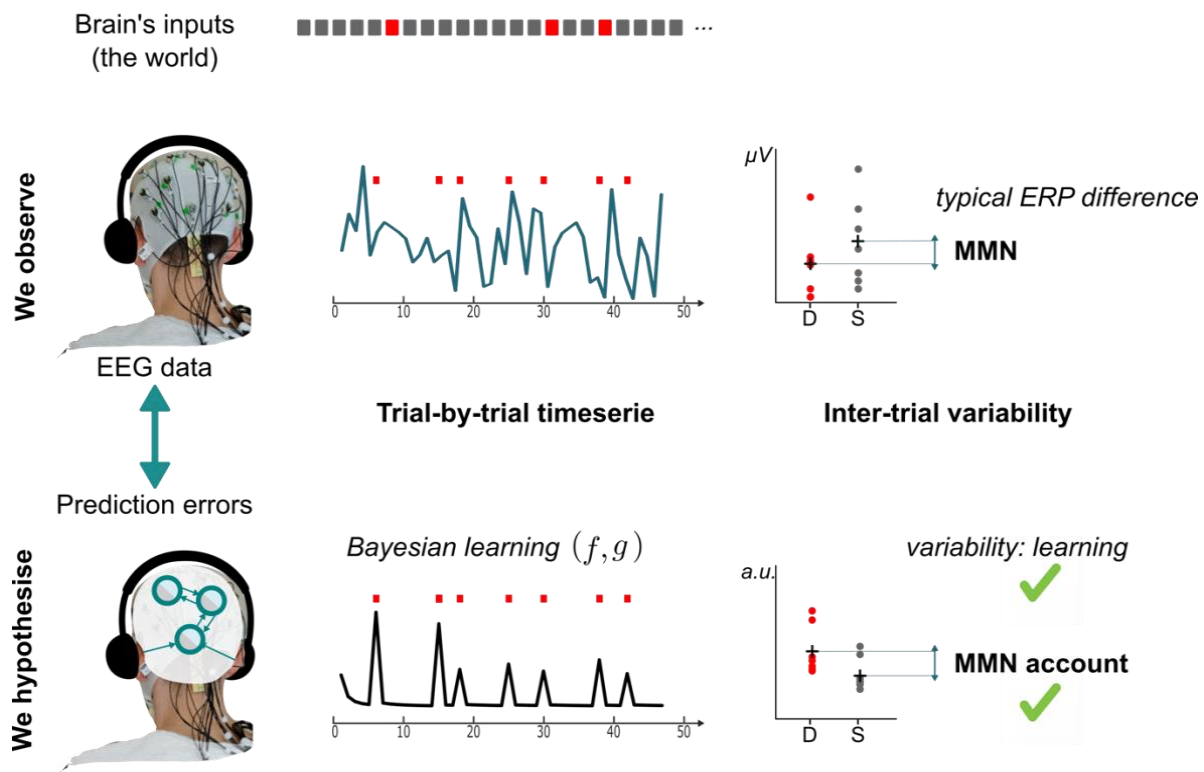

## b) Static models ignore variability (as in ERP/ERFs)

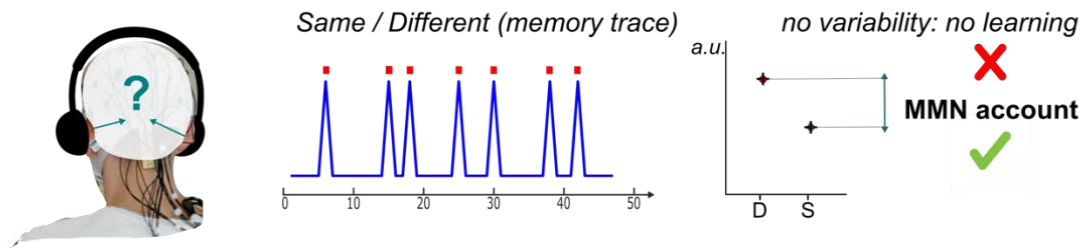

*Figure 6. a) The PC model of the MMN and oddball processing. Assuming that the brain aims at inferring the generative model of the world, it could learn the oddball rule when exposed to an oddball stream. Upper panel: observations by the experimenter illustrated using EEG data. A trial-by-trial timeseries is shown (electrode Fz, trial 1 to 50, t = 150 ms) using the same display as in Figure 5. Right: signal amplitudes (50 samples) are sorted by stimulus category (D:deviant, S: standard preceding a deviant). The difference between deviant and standard averages (black dots) reflects an MMN. Lower panel: PC model predictions. The brain entertains the Bayesian learning of the oddball rule. A possible hypothesis (with f, g, described in Lecaignard et al., 2022a) yields a trajectory of prediction errors as Bayesian surprise over trials (black trace). Right: same display as with EEG data. This model succeeds in producing a deviant-standard difference. **b) The memory-based model of the MMN.** This model predicts the same error signal for every deviant (blue trace). This model lacks the mechanistic description providing the memory trace (f, g not explicit). The absence of trial order effect is the signature*

*of non-learning (past experience does not matter). Important, although not supported by empirical data (the MMN is context-sensitive), this model predicts an MMN.*